\documentstyle[prb,aps,eqsecnum]{revtex}  

\begin{document}

\title{Implications of a minimum in the temperature dependent electrical
resistance above the magnetic ordering temperature in  Gd$_2$PdSi$_3$ }
\author{R. Mallik and E.V. Sampathkumaran$^{*}$ }
\address{Tata  Institute  of  Fundamental  Research,  Homi   Bhabha
Road,   Colaba, Mumbai-400 005, INDIA }
\author{M. Strecker and G. Wortmann}
\address{Fachbereich 6-Physik, Universit\"at Paderborn, D-33095, GERMANY} 
\maketitle

\baselineskip 0.5cm
\vskip 1.5cm
Ever since  a  minimum  in  the   electrical   resistance   (R)   at   a
characteristic temperature in noble metals containing  3d  magnetic
impurities was reported several decades ago,  the consequences  of  this
phenomenon, known as the Kondo effect, in metallic solids remains  an
active area of research. In concentrated Kondo alloys of Ce, Sm, Yb or U
[1], if the strength of the intersite  indirect exchange  interaction
(given  by $T_{RKKY})$ is comparable to that of the Kondo effect (given by
the Kondo temperature, $T_{K})$, one also observes magnetic ordering well
below the temperature  at which resistivity  shows  a minimum.  However,
such a feature in R is not expected for Gd alloys, since the Gd-4f orbital
is so well localized that it cannot exhibit the Kondo effect.  In this
article, we report the observation of a pronounced minimum in the plot of
R versus temperature (T) well above the magnetic ordering temperature in a
Gd based intermetallic compound, Gd$_2$PdSi$_3$ (which is presumably
antiferromagnetic [2], T$_N$= 21 K), resembling the behaviour in magnetic
Kondo lattices. This finding implies that a minimum in R(T) can also occur
from a completely different mechanism as a magnetic precursor effect, which,
we believe, is an exchange interaction induced electron localization.
\par
The  polycrystalline sample, Gd$_{2}$PdSi$_{3}$,  was  prepared  by  arc
melting followed by homogenization at  750$^o$C  in an evacuated sealed
quartz tube.  We   have   also investigated  alloys with Gd substituted by
Y, i.e., (Gd$_{1-x}$Y$_{x})_{2}$PdSi$_{3}$ (x= 0.2, 0.5 and 0.8). The  x-
ray powder diffraction patterns (Cu  $K_{\alpha }$) confirm that these
alloys are single phase forming the hexagonal AlB$_{2}$-type structure
[2]. The homogeneity of the samples were checked by scanning electron
microscopy. The  electrical resistance measurements  (4.2 - 300 K) were
performed  by  a conventional four-probe method.  Additional experiments
carried out to support the line of our arguments and conclusions are:  (i)
The magnetoresistance measurements in the longitudinal geometry in a
magnetic field (H) of 50 kOe as a function of T and  as a function of  H
at selected temperatures for the $x$= 0.0 alloy; (ii) $^{155}$Gd (5/2
$\rightarrow$ 3/2, 86.5 keV transition) M\"ossbauer measurements in the
transmission geometry with a 2 mCi $^{155}$Eu(SmPd$_3$) source for $x$=
0.0 alloy below 25 K at selected temperatures; (iii) Heat-capacity (C)
measurements as a function of temperature  (2 - 70 K); (iv) Magnetic
susceptibility ($\chi$) measurements (2 - 300 K) in  a field of 100 Oe and
in 2 kOe.
\par
The temperature dependence of the normalized electrical resistivity below
200 K  is  plotted in figure 1 for all compositions investigated. The
values of $\rho $ at  300 K  and 4.2 K  for Gd$_{2}$PdSi$_{3}$ are of the
order of 400  and 280 $\mu$$\Omega$ cm respectively. Due to the presence
of microcracks in the sample, the error of these values might  be smaller
by about 40\%.  For this reason, the data shown in Fig. 1 are normalized
to  the values at 300 K. It is obvious from Fig.  1 that the $\rho$(T) of
the Gd$_{2}$PdSi$_{3}$ compound gradually decreases as the temperature is
lowered down to 60 K, followed by an upturn below about  45 K  thus giving
rise to a resistivity minimum at about (T$_{min}$=) 45 K; the value before
the onset of magnetic ordering (say, at 22 K)   is about  5\% higher
compared to the minimum resistivity.  There is  a  kink in $\rho$(T)
versus T at about  20 K, close to the temperature where the Gd sublattice
exhibits  long  range magnetic ordering presumably of
antiferromagnetic-type, as one may conclude from the well-defined peak  in
the magnetic susceptibility and from the shoulder in the heat  capacity
data ({\it vide infra}).  Further lowering of temperature does not result
in a drop in $\rho$(T) as expected due  to  the loss of the spin-disorder
contribution below the N\'eel temperature (T$_N$).  We presently attribute
this to the opening of a pseudogap  at some portions of the Fermi surface
due to the onset of antiferromagnetism in this compound [3]. The  magnetic
ordering temperatures  for the compositions  $x$= 0.2, 0.5  and 0.8 are
lowered (see the heat-capacity data described below) as expected on the
basis of indirect exchange mechanism; however, the resistivity minimum
persists well above T$_N$ for all compositions. Even in the diluted limit
of our investigation,  $x$= 0.8, the resistance increases by about 2\%  as
the temperature is lowered from T$_{min}$ towards 4.2 K. The fact that the
minimum is also observable in the sample with the highest Y content (x=
0.8) may exclude the explanation based on the opening of an energy gap
(above 21 K).
\par
Now turning to the influence of the application of a magnetic field,  the
negative $d\rho /dT$ behavior in zero field vanishes in the presence of a
field of 50 kOe (shown in Fig.  2a for $x$= 0.0 only for the sake of
clarity). The exact dependence of T$_N$ on  H  could not be inferred from
the present data and it is not  relevant  to  the main conclusion of this
article. The magnetoresistance, defined  as $\Delta \rho /\rho  =
\{\rho (H)-\rho (0)\}/\rho (0)$, is obtained from this data as a function
of  T (shown in Fig.  2b for all the compositions).  For $x$= 0.0, it is
found that $\Delta \rho /\rho $ is as large as about  -10\%  at  40 K  in
the presence of  H= 70  kOe (Fig. 2c). For comparison, the magnitude of
the magnetoresistance values for the isostructural  Ce based Kondo lattice
Ce$_{2}$PdSi$_{3}$ (Ref.  4)  as well as for the alloys,
(Ce$_{1-x}$Y$_{x})_{2}$PdSi$_{3}$, is less than 1\%  down to  4.2 K. This
establishes that the observed large magnetoresistance effect is
Gd-related, as noted by us for some other Gd alloys [5].
\par
It is not out of place to add that, as the temperature is lowered below
T$_N$, the $\rho $(T) curve at 50 kOe does not show an upturn.  If the
upturn in $\rho$ below T$_N$ is attributed to magnetic Brillouin-zone
boundary gaps, this finding implies that these gaps are essentially
getting washed out by the application of a magnetic field; there is a
feeble  rise below  7 K in the data in the presence of a field, due to
possible non-vanishing gap   at low temperatures.  The magnitude of
$\Delta\rho /\rho $ thus increases below 20 K for the $x$= 0.0 alloy (Fig.
2b), attaining a  peak value  of about -32\%  at 7 K.  In order to
highlight this feature further, the $\Delta\rho /\rho $ is plotted as a
function of H  in Fig.  2c  for various temperatures. The overall shape of
the plot of $\Delta\rho /\rho $ versus H   gets significantly modified as
the temperature  is lowered across T$_N$ due to suppression of the
proposed magnetic Brillouin zone boundary gap.  The observation of a large
negative value  of $\Delta \rho /\rho $ is of importance considering the
current interest  in the phenomenon of giant magnetoresistance.
\par
In order to be sure that the upturn in R below T$_{min}$ is not due to the
opening up of magnetic Brillouin-zone boundary gaps, it is desirable to
confirm the T$_N$ value by a microscopic technique as well. We therefore
show the results of $^{155}$Gd M\"ossbauer (only for $x$= 0.0), $\chi$ and
C measurements in Fig. 3, 4a and 4b respectively.  The M\"ossbauer spectra
below 20 K are found to undergo Zeeman splitting, which is well-resolved
below 15 K. The magnetic hyperfine field at the Gd nucleus obtained by a
standard least square fitting procedure of the M\"ossbauer spectra are
also shown in figure 3 for $x$= 0.0.   The temperature at which the
magnetic hyperfine field extrapolates to zero (near 20 K) marks the
magnetic ordering temperature and this serves as a conclusive microscopic
proof for paramagnetism above 20 K (which is also confirmed by our
M\"ossbauer thermal scan experiment) in this compound. The spectral
analysis of the data for 25 K gave evidence for a quadrupolar splitting
only (the magnitude of which is 0.88mm/s). 
\par
We have also probed the magnetic behavior by other bulk measurements. With
respect to $\chi$ (H= 2 kOe), the low temperature data show a distinct
cusp in $\chi$ at about 20 K in this alloy, indicative of
antiferromagnetic-like ordering.  The feature due to magnetic ordering
shifts to lower temperatures with dilution of the Gd sublattice (Fig.
4a).  The plot of the $\chi$$^{-1}$ versus temperature (not shown in a
figure) is found to be linear down to the critical temperature, thus
suggesting that the peak in $\chi$(T) marks the onset of long range
magnetic order.  Consistent with this, in the heat capacity there is a
peak attributable to the onset of magnetic ordering at different
temperatures (Fig.  4b) for each of  the three compositions, $x$= 0.0, 0.2
and 0.5 with a magnetic ordering temperatures of 20, 15 and 8 K
respectively. For $x$= 0.8, there is no clearcut anomaly due to the onset
of long range magnetic ordering above 2 K. We obtained the 4f contribution
$(C_{m})$  to  C, shown in Fig. 4c, using the procedure suggested in Ref.
6 employing the  C values of Y$_{2}$PdSi$_{3}$ as a reference for the
lattice contribution.  A point to be noted [7] is that there  is  a broad
shoulder (marked by a vertical arrow in Fig.  4c) at about  22 K (which is
a  measure of T$_N$ from heat capacity [6]) for $x$= 0.0, beyond which
$C_{m}$  falls gradually with temperature, extending to about 20 K above
T$_N$, similar to that in other  Gd alloys [8]. The magnetic entropy
$(S_{m})$  at T$_N$ obtained from our data is only about 75\% of the
saturation value of R$\ln$8 (R= gas-constant) for all the
magnetically ordered alloys;  the full  value is reached far above T$_N$
only, for instance, only around  45 K  in Gd$_{2}$PdSi$_{3}$.  Clearly,
the heat-capacity tail persists for about 20 K due to magnetic precursor
effects [8] even in the case of  the dilute  alloy, $x$= 0.8, mimicking
heavy-fermion behavior, though this alloy does not undergo long range
magnetic order above 4 K.
\par
Thus, all the results, including those from the M\"ossbauer spectroscopy,
conclusively establish the onset of long range magnetic ordering  in the
vicinity of 21 K for $x$= 0.0. Hence the upturn in R below 45 K is not a
consequence of a long range antiferromagnetic ordering (gap) effect.  It
is important to note that the temperature till which the heat-capacity
tail  above T$_N$ persists and the one below which a negative $\Delta\rho
/\rho$ is noticeable beyond experimental error shift down with increasing
$x$ (compare the data for $x$= 0.0 and 0.8 in Fig.  2b) [9].  These
observations suggest that these anomalies arise from a common physical
origin and thus the minimum in R is a manifestation of a Gd 4f magnetic
precursor effect.  We would like to add that there is no difference
between the field-cooled and zero-field-cooled susceptibility (H= 100 Oe)
above T$_N$ in these alloys and hence spin-glass effects need not be
considered in the temperature region of interest.
\par
It should be mentioned that the temperature dependent x-ray diffraction
studies do not show any structural change down to 12 K; no anomaly is
apparent in the lattice constants across T$_N$ as well [10]. This rules
out structural anomalies in the vicinity of T$_N$ as a cause of this R
minimum.  It is not clear whether anomalous critical scattering effects or
spin fluctuations involving Pd 4d [11] and/or Gd 5d conduction electrons
are responsible for the minimum in R(T). The spin fluctuations cannot be
of the conventional Kondo type, as the  4f level of Gd in general is
believed to be situated well below  the Fermi level.  It may be added that
reports confirming Pd 4d ion Kondo effect exist only in the single-ion
limit [12]; if the effect is caused by Pd 4d, it is clearly an effect
induced by Gd, since no  R minimum has been observed in Y$_2$PdSi$_3$ [4].
Above all, the sign of the Curie-Weiss temperature obtained from the high
temperature $\chi$ data is not negative [2] as it should be negative for
conventional Kondo systems. Therefore these spin fluctuations, if they exist,
are triggered by Gd 4f short range correlations {\it prior to long range
magnetic ordering}. We propose that the origin of the R(T) minimum lies in
a {\em localization } or trapping of the electron cloud polarized by the
s-f (or d-f) exchange interaction as one lowers the temperature towards
long range magnetic transition temperature; this could be triggered by Gd
4f short range magnetic order [8], just as crystallographic disorder
results in electron localization [13], thereby reducing the mobility of
charge carriers. The application of a magnetic field suppresses the  spin
fluctuations or alters the localization length (as proposed for La
manganites [14] above the Curie temperature) resulting in an enhancement
of the mobility of the electron cloud; this explains the suppression of
the R minimum in the paramagnetic state in  a magnetic field.  The
concepts of exchange-interaction mediated localization, i.e., some kind of
magnetic-polaronic effects, have been proposed for  semiconducting
rare-earth compounds before [15].  Through this article we point out that
the essential concept discussed in Refs. 14 and 15 may to some extent be
applicable even in some metallic alloys, particularly to those behaving as
poor metals, i.e., in which the residual resistivity ratio is low, as it
appears to be the case in Gd$_2$PdSi$_3$.
\par
Summarising,  the intermetallic compound, Gd$_2$PdSi$_3$, is  shown  to
exhibit an  electrical  resistivity  minimum  as  a  function of
temperature above T$_N$.  We propose that a novel magnetic precursor
effect, possibly electron localization {\it the rootcause of which lies in
magnetism even in metallic systems}, may result in an upturn of the
electrical resistance before long range order sets in. Finally, we have
also made several other interesting observations in this compound. These
are:  (i) Positive Curie-Weiss temperature indicative of canted
antiferromagnetism; (ii) The absence of downturn in electrical resistance
below T$_N$; (iii) A sudden jump in the $\mid$B$_{eff}\mid$ at the Gd
nucleus at 15 K, related to lower ordering temperature for one of the two
Gd sites [2,4] and (iv) Negative magnetoresistance with the magnitude
increasing with decreasing temperature (large near T$_N$), an observation
of significance to the field of giant magnetoresistance.

\vskip 1cm
\centerline{REFERENCES}
\vskip 0.5cm
\noindent [1] M.B. Maple, L.E. DeLong and B.C. Sales, in "Handbook  on the
Physics and Chemistry  of  Rare earths,"  ed.,  K.A.  Gschneidner, Jr.
and  L.  Eyring (North-Holland, Amsterdam, 1978) 797 and references
therein.
\newline [2] P.A. Kotsanidis, J.K. Yakinthos and E. Gamari-seale, J. Magn.
Magn.  Mater.  {\bf 87} (1990) 199.
\newline [3] R.J. Elliot and F.A. Wedgewood, Proc. Phys. Soc. {\bf81}
(1963) 846; I. Das, E.V. Sampathkumaran and R. Vijayaraghavan, Phys.
Rev.  B {\bf 44} (1991) 159  and references therein.
\newline [4] R. Mallik and E.V. Sampathkumaran, J. Magn. Magn. Mater.
{\bf 164} (1996) L13.  
\newline [5] I. Das and  E.V.  Sampathkumaran,  Phys.  Rev.  B {\bf 49} (1994)
3972;  E.V.  Sampathkumaran and I. Das, Phys. Rev.  B {\bf 51} (1995)
8631;  Physica  B {\bf 223\&224} (1996) 149; R. Mallik et al., Phys. Rev.
B {\bf 55} (1997) R8650.
\newline [6] J.A. Blanco, P. Leuthuilier, and D. Schmitt, Phys. Rev.  B
{\bf 43} (1991) 13137. 
\newline [7] The peak values in Fig. 4c (close to 10 mJ/Gd mol), are far
less than  that  expected [6] (27 J/ Gd mol) for  an equal-moment
antiferromagnetic structure and this  proves that the magnetic  structure
could be of modulated or of any other complicated type. We retain the
notation T$_N$ to represent the onset of magnetic ordering. 
\newline [8] E.V. Sampathkumaran and I. Das, Phys. Rev.  B {\bf 51}
(1995) 8178.
\newline [9] The value of T$_{min}$ is found to be specimen-dependent
(possibly due to slight non-uniform distribution of Gd in the dilute
limit) for the composition, x= 0.8, and it falls in the range 15-45 K.
Therefore, the magnetoresistance data (Fig. 2b) is more instructive to
infer the onset temperature of magnetic precursor effect for this alloy.
\newline [10] R. Mallik, E.V. Sampathkumaran, M. Strecker, G. Wortmann,
P.L. Paulose and Y. Ueda, Phys. Rev. B (submitted). 
\newline [11] See, for instance, E. Gratz, R. Resel, A.T. Burkov, E.
Bauer, A.S. Markosyan and A. Galatanu, J. Phys.: Condens. Matter. {\bf7}
(1995) 6687. 
\newline [12] K.D. Gross, D.  Riegel and R.  Zella,  Phys.  Rev. Lett.
{\bf 65} (1990) 3044.
\newline [13] P.A. Lee and T.V. Ramakrishnan, Rev. Mod. Phys. {\bf57}
(1985) 287.
\newline [14] C.M. Varma, Phys. Rev. B {\bf54} (1997) 7328; M. Viret, L.
Ranno, and J.M.D. Coey, Phys. Rev. B {\bf55} (1997) 8067.
\newline [15] T. Kasuya, J. Appl. Phys. {\bf77} (1995) 3200; Y. Shapira,
S. Foner, N.F. Oliveira,Jr and T.B. Reed, Phys.  Rev. B {\bf10} (1974)
4765; P. Nyhus, S. Yoon, M. Kauffman, S.L. Cooper, Z. Fisk and J. Sarrow,
Phys. Rev. B {\bf56} (1997) 2717.

\begin{figure}
\caption{
The electrical resistance (R) of the alloys, (Gd$_{1-x}$Y$_{x})_{2}$PdSi$_{3}$
($x$= 0.0, 0.2, 0.5  and 0.8), normalized to 300 K  values.  }
\end{figure}

\begin{figure}
\caption{
(a)  The electrical  resistivity $(\rho )$  as  a  function  of
temperature  for Gd$_{2}$PdSi$_{3}$ in the presence and in the absence of
a magnetic field of 50 kOe.  (b)  The magnetoresistance $(\Delta \rho
/\rho )$ obtained from the data shown in (a) for $x$= 0.0 is plotted; in
addition, the data obtained in a similar way for other compositions are
also shown.  (c) For $x$= 0.0, the $\Delta \rho /\rho
$ as a function of magnetic field at selected temperatures;  the lines
through the data points serve as guides to the eye. The error in $\Delta
\rho /\rho $ is estimated to be less than 0.5\% .  }
\end{figure}
\begin{figure}
\caption{
$^{155}$Gd M\"ossbauer spectra of Gd$_2$PdSi$_3$ at selected temperatures.
The continuous lines represent standard least square fit of the data. At
the bottom, the effective magnetic hyperfine field values obtained from
the data are shown as a function of temperature.}
\end{figure}
\begin{figure}
\caption{
(a) The magnetic susceptibility  (H= 2 kOe) as a function of temperature,
(b) the heat capacity and (c)  The  4f   contribution $(C_{m})$  to C as
a function of  temperature  for  the  alloys,
(Gd$_{1-x}$Y$_{x})_{2}$PdSi$_{3}$. The lines through the data points in
(a) serve as guides to the eyes.}

\end{figure}

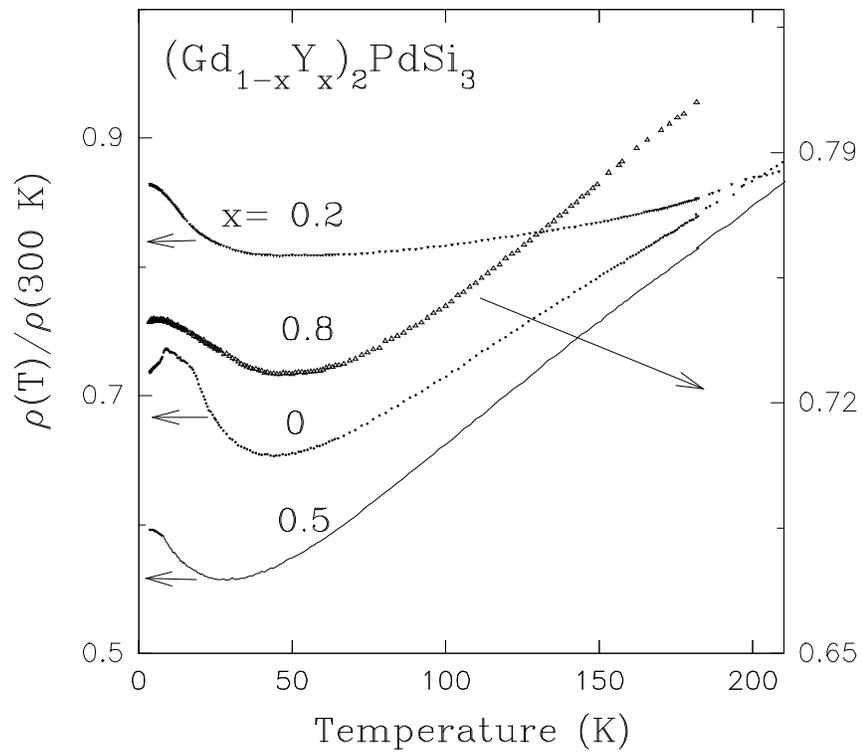
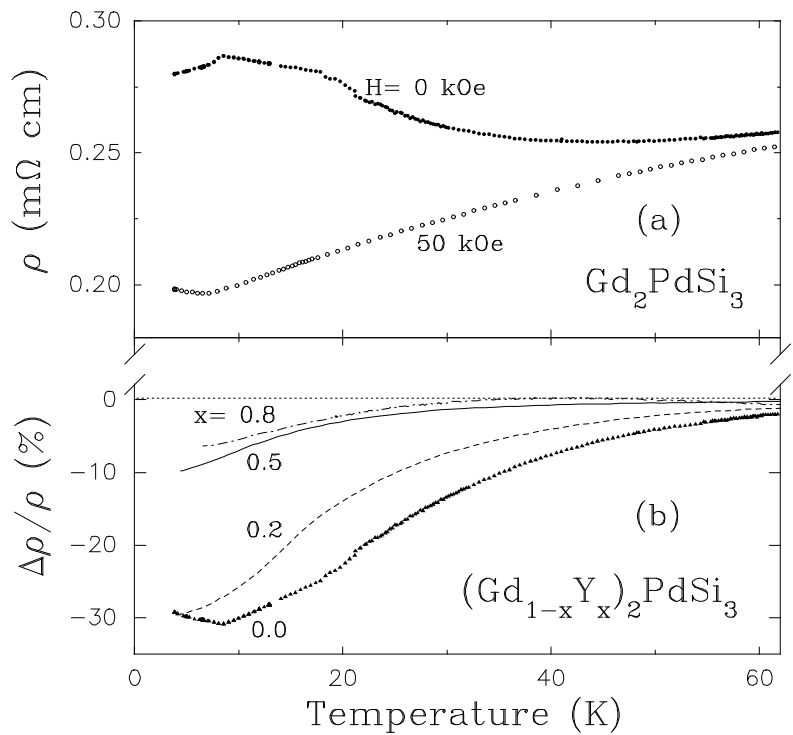
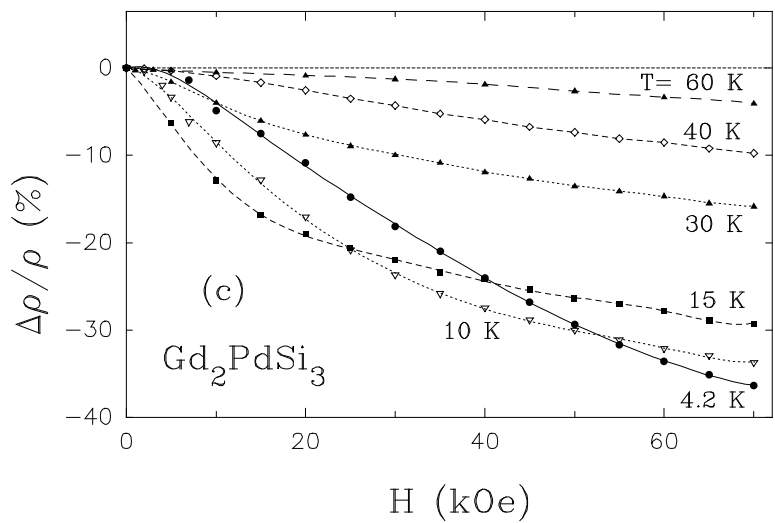
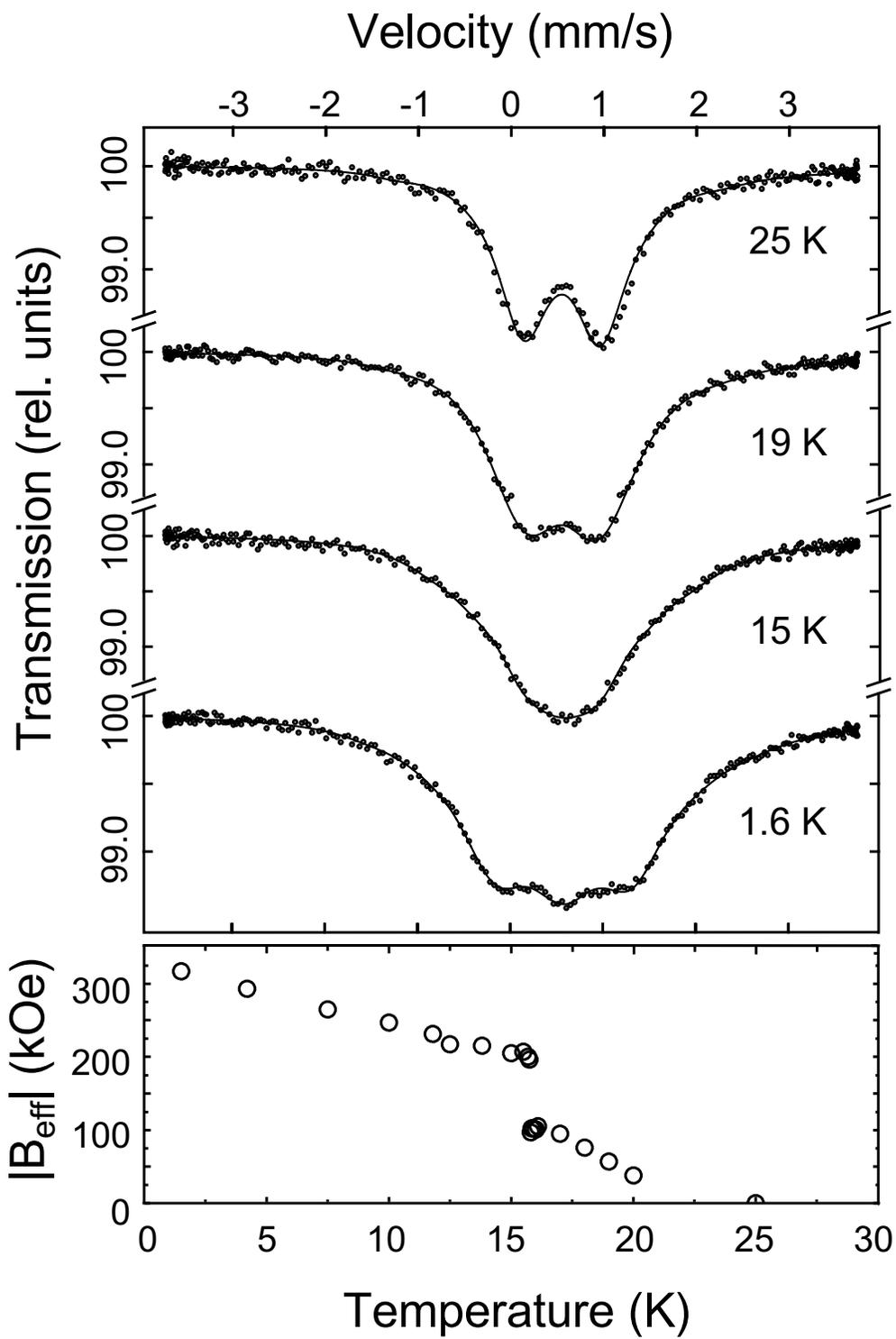
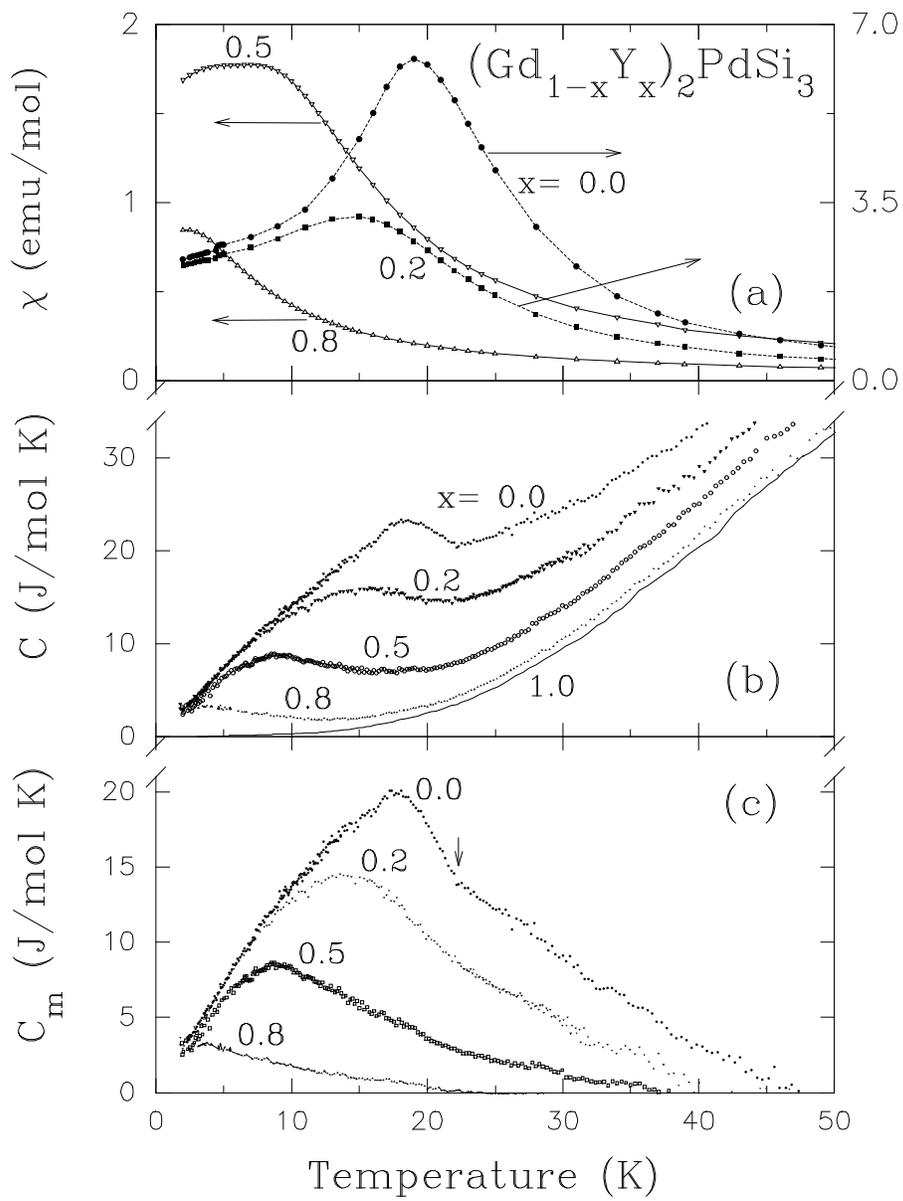
\end{document}